# High-$T_C$ superconductivity in $Cs_3C_{60}$ compounds governed by local Cs-$C_{60}$ Coulomb interactions

**Dale R. Harshman[1] and Anthony T. Fiory[2]**

[1] Department of Physics, The College of William and Mary, Williamsburg, Virginia 23187, USA
[2] Department of Physics, New Jersey Institute of Technology, Newark, New Jersey 07102, USA

E-mail: drh@physikon.net

**Abstract**

Unique among alkali-doped $A_3C_{60}$ fullerene compounds, the A15 and fcc forms of $Cs_3C_{60}$ exhibit superconducting states varying under hydrostatic pressure with highest transition temperatures at $T_C^{meas}$ = 38.3 and 35.2 K, respectively.  Herein it is argued that these two compounds under pressure represent the optimal materials of the $A_3C_{60}$ family, and that the $C_{60}$-associated superconductivity is mediated through Coulombic interactions with charges on the alkalis.  A derivation of the interlayer Coulombic pairing model of high-$T_C$ superconductivity employing non-planar geometry is introduced, generalizing the picture of two interacting layers to an interaction between charge reservoirs located on the $C_{60}$ and alkali ions.  The optimal transition temperature follows the algebraic expression, $T_{C0}$ = (12.474 nm$^2$ K)/$\ell\zeta$, where $\ell$ relates to the mean spacing between interacting surface charges on the $C_{60}$ and $\zeta$ is the average radial distance between the $C_{60}$ surface and the neighboring Cs ions.  Values of $T_{C0}$ for the measured cation stoichiometries of $Cs_{3-x}C_{60}$ with x ≈ 0 are found to be 38.19 and 36.88 K for the A15 and fcc forms, respectively, with the dichotomy in transition temperature reflecting the larger $\zeta$ and structural disorder in the fcc form.  In the A15 form, modeled interacting charges and Coulomb potential $e^2/\zeta$ are shown to agree quantitatively with findings from nuclear-spin relaxation and mid-infrared optical conductivity.  In the fcc form, suppression of $T_C^{meas}$ below $T_{C0}$ is ascribed to native structural disorder.  Phononic effects in conjunction with Coulombic pairing are discussed.

# 1. Introduction

The buckminsterfullerene, i.e. the $C_{60}$ molecule, was the first fullerene to be successfully synthesized [1], with the production of significant quantities of $C_{60}$ crystals subsequently reported [2]. Following the discovery of metallic conduction in alkali-doped $C_{60}$ [3], superconductivity was reported for face-centered cubic (fcc) $A_3C_{60}$ compounds at ambient pressure with alkali $A$ (as single or dual species) occupying the two tetrahedral (T) and one octahedral (O) interstitial sites per $C_{60}$ [4-6].

The highest transition temperatures for the $A_3C_{60}$ family are found in the $Cs_3C_{60}$ materials: $T_C^{meas} = 38.3$ K for the more ordered A15 structure with bcc $C_{60}$ packing (Pm3n) and $T_C^{meas} = 35.2$ K for fcc packing (Fm3m), as measured under hydrostatic pressures of 0.93 and 0.73 GPa, respectively [7,8], and corresponding to diamagnetic shielding fractions approaching maxima [8,9] (see also [10] for earlier work on A15 $Cs_3C_{60}$). Both macrostructural forms also exhibit an antiferromagnetic (Mott) insulating state at zero applied pressure [7,8]. An important difference between these two forms exists in the ordering of the $C_{60}$ molecules; the merohedral (orientational) disorder of the $C_{60}$ molecules evident in the fcc $A_3C_{60}$ superconductors [11,12], including fcc $Cs_3C_{60}$, is absent in A15 $Cs_3C_{60}$ [7,9].

In a recent study of superconducting $Rb_xCs_{3-x}C_{60}$ under pressure, dome behavior, i.e. a local maximum, was observed in the variation of $T_C$ with the volume per $C_{60}$ ion ($V_{per\ C_{60}}$) for $x \leq 1$, where maximum transition temperature $T_C^{max}$ and corresponding $V_{per\ C_{60}}$ increase as $x \rightarrow 0$ [13]. The same can be stated for A15 $Cs_3C_{60}$, which also exhibits a dome-like dependence of $T_C$ on $V_{per\ C_{60}}$ [7,8]. These results demonstrate that A15 and fcc $Cs_3C_{60}$ under hydrostatic pressure represent the optimal and nearly optimal compounds for these two structural forms, respectively, wherein $T_C^{meas}$ is taken to equal $T_C^{max}$; the presence of $C_{60}$-merohedral and Cs(O)-site disorder in fcc $Cs_3C_{60}$ (and fcc $A_3C_{60}$ in general) [11,12,14] is expected to suppress $T_C^{meas}$.

Correlation between $T_C$ and lattice parameter $a_0$ (equivalently, $V_{per\ C_{60}}$) for fcc $A_3C_{60}$ compounds at ambient pressure [6,15] has been cited as evidence of phonon involvement, possibly indicative of a BCS (Bardeen, Cooper, Schrieffer) type of mechanism wherein the $C_{60}$–$C_{60}$ separation determines the electron density of states at the Fermi level (DOS). This approach typically assumes that the intramolecular modes (34–195 meV, see e.g. [16]) play the dominate role in superconductivity, and focuses on electron-phonon coupling constants in the presence of an enhanced Jahn-Teller effect. Seemingly not understood from theoretical treatments along these lines are the depressed $T_C$ and vanishing superconductivity observed for the compounds $Li_xCsC_{60}$ [17], $Na_2Cs_xC_{60}$ [18] and $K_{3-x}Ba_xC_{60}$ [18] with the $C_{60}$ charge state deviating from the nominal −3 (e.g. by ± 0.5 for integer $x$) [19], particularly in view of the theoretically expected smooth DOS near half filling of the $t_{1u}$ band [18,20]. An optimal doping level is evidently fundamental in $A_3C_{60}$ and evocative of high-$T_C$ superconductivity (e.g. doping in cuprates [21]). Early theoretical works also do not anticipate the non-linear and non-monotonic behavior in $T_C$ vs. $V_{per\ C_{60}}$ subsequently reported for $Rb_xCs_{3-x}C_{60}$ under pressure, although a possibly maximum $T_C$ was mentioned (an extensive review is given in [22]). A more recently presented theoretical $T$-$V_{per\ C_{60}}$ phase diagram for $Cs_3C_{60}$ was derived from a Hubbard model and negative Hund's coupling, finding a maximum $T_C$ at the metal-insulator phase boundary [23] (methodology reviewed in [24,25]), the locus of which follows results from $Rb_xCs_{3-x}C_{60}$ [13].

The optimization behavior observed as maxima in the variations of $T_C$ with $V_{per\ C_{60}}$ for A15 [7] and fcc [8] $Cs_3C_{60}$ is reminiscent of characteristics typically found in high-$T_C$ materials. Telling evidence of unconventional superconductivity in $Cs_3C_{60}$ is the absence of a Hebel-Slichter coherence peak in the nuclear spin relaxation rates in the A15 form under pressure [26]. Additionally, for fcc $RbCs_2C_{60}$, where $T_C$ = 32.9 K is nearly maximized at ambient pressure ($T_C^{max}$ = 33.2 K at $P \approx 0.17$ GPa), both the specific heat jump and the superconducting energy gap notably exceed weak-coupling BCS expectations [13]. Pursuing the analogy to high-$T_C$ in this work, $T_C^{meas}$ is identified with $T_{C0}$ determined by the interfacial Coulomb-mediated pairing model in [27]. Data for other $A_3C_{60}$ superconductors at ambient pressure provide a continuation of the $T_C$ variation to lower values (e.g. figure 4(a) in [8]), together with a diminishing Meissner effect [6] and apparent superfluid density [28],[1] in a manner consistent with non-optimal behavior. Owing to merohedral disorder and large spatial fluctuations, particularly of the Cs ions occupying the (O) sites [12,14], fcc $Cs_3C_{60}$ provides a good test case for the role of such defects in suppressing $T_C^{meas}$ below $T_{C0}$. Consequently, the present work focuses on the optimal $T_C$ regions of the $T$-$V_{per\ C_{60}}$ phase diagrams for A15 and fcc $Cs_3C_{60}$, where it is found that interfacial Coulombic interactions dominate.

The Coulombic-based model [27] discussed herein locates the interacting charges in two reservoirs, one superconducting (type I) and the other mediating (type II). In the layered high-$T_C$ superconductors, the two reservoirs are formed in adjacent layered-crystal structures separated by an interaction distance $\zeta$. Thin film studies of cuprates have determined that one each of the two charge reservoirs is sufficient to create and sustain the high-$T_C$ superconductive state [29]. When viewing the macroscopically cubic $A_3C_{60}$ packing structures from a local perspective, the C sites on a given $C_{60}$ can be treated as a single layer with non-planar character. Since superconducting pairs form locally, the $C_{60}$ is identified as the type I structure and the surrounding nearest-neighbor alkalis as the type II structure. Within the Coulombic pairing model, the charges on $C_{60}$ anions interact with the charges associated with alkali cations, distinguishing it from theories founded on intramolecular coupling. The presence of cationic charges is indicated from NMR spin-lattice relaxation, hyperfine coupling constants and other measurements, which are discussed in section 3 [12,26].

Section 2 presents the derivation adapted to $Cs_3C_{60}$ of the Coulombic pairing model based on the inter-reservoir Coulomb interaction between physically separated charge interfaces; $T_{C0}$ values for the optimal A15 and fcc forms of $Cs_3C_{60}$ (under optimal applied hydrostatic pressure) are calculated. Key experimental results and interpretations are considered in section 3 and conclusions are summarized in section 4.

## 2. Interlayer Coulombic Pairing Model

The model first described in [27] regarding the pairing mechanism governing high-$T_C$ superconductivity assumes a layered 2D-like interaction structure comprising a superconducting type I charge reservoir and a mediating type II reservoir, typically of opposite sign, that are physically separated by an interaction distance $\zeta$ defined normal to the layers. In the case of $YBa_2Cu_3O_{7-\delta}$, for example, the interaction occurs between adjacent BaO ($p$-type) and $CuO_2$ ($n$-type) 2D layers, separated by an interaction distance $\zeta$, with the former designated as part of the type I reservoir (BaO-CuO-BaO) and the latter assigned to the type II reservoir ($CuO_2$-Y-$CuO_2$). Consideration of thin-film samples of $YBa_2Cu_3O_{7-\delta}$, $Bi_2Sr_2CaCu_2O_{8+\delta}$ and $La_{2-x}Sr_xCuO_4$ has, in fact, shown that the minimum superconducting entity is commensurate with a single formula unit structure [29]. The presence of these two disparate charge reservoirs is the probable source of differing conclusions regarding the ground-state symmetry in some high-$T_C$ superconductors [30]. At the time of this writing, this model has already been validated with a statistical deviation between the calculated and measured $T_{C0}$ of ±1.35 K for 48 different layered materials from seven superconducting families (cuprates, ruthenates, rutheno-cuprates, iron-pnictides and ET-based [bis(ethylenedithio)tetrathiafulvalene] organics [27,31,32]; iron-chalcogenides [33]; intercalated group-4-metal nitride-halides [34,35]) with measured $T_C^{meas}$ values ranging from ~7 to 150 K.

---

[1] Muon-spin depolarization rates for ambient pressure $A_3C_{60}$ superconductors resemble those for non-optimally doped cuprate superconductors.

## 2.1. $T_{C0}$ and charge allocation

As originally formulated for layered structures, the algebraic expression defining the optimal transition temperature $T_{C0}$ is given by [27],

$$T_{C0} = k_B^{-1} \beta (\sigma\eta/A)^{1/2} \zeta^{-1} = k_B^{-1} \beta (\ell\zeta)^{-1} . \quad (1)$$

Here, $\sigma$ is the fractional charge for participating carriers per formula unit, $A$ is the basal plane area, which is the same for the two reservoirs in layered structures, $\eta$ is the number of charge-carrying type II layers (e.g. $\eta = 2$ for $YBa_2Cu_3O_{7-\delta}$), and the universal constant $\beta =$ 0.1075 ± 0.0003 eV Å$^2$ was determined previously from experimental data for $T_C^{meas}$ [27]. The length $\ell = (A/\sigma\eta)^{1/2}$ defined in equation (1) relates to the mean spacing between interacting charges. The $T_{C0}$ defined in equation (1) should be considered as an upper limit on the experimentally observed transition temperature, given $T_C < T_{C0}$ for non-optimal materials. The optimization of the superconducting state is achieved when the two interacting charge reservoirs are in equilibrium.

Defining $\beta = e^2\Lambda$, where $\Lambda = 0.00747$ Å is about twice the reduced electron Compton wavelength, points out the presence of the Coulomb potential $e^2\zeta^{-1}$ in equation (1). As such, the physics contained in equation (1) is interpreted in terms of superconductive pairing mediated by coupling or exchange of virtual bosons with energies on the order of $e^2\zeta^{-1}$ [27]. A related model with spatially indirect Coulomb interactions involving charges on neighboring ions within the unit cell was first suggested in 1987 as a possible electronic excitation mechanism for high-$T_C$ superconductivity [36]. Around the same time it was also proposed that interactions between neighboring cations and anions are unscreened, owing to the low free electron density in high-$T_C$ materials [37]. Not surprisingly, high energy components of electron-boson coupling functions for high-$T_C$ cuprate superconductors are, in fact, experimentally observable in thermal reflectance spectra [38,39].

Direct doping may be either cationic or anionic, occurring in the type I reservoir as in the case of $La_{2-x}Sr_xCuO_{4-\delta}$, the type II reservoir, e.g. $Ba_2Y(Ru_{1-x}Cu_x)O_6$ [27], or both as in the ternary Fe-based chalcogenides (e.g. $A_xFe_{2-y}Se_2$ [33]) and $(Ca_xLa_{1-x})(Ba_{1.75-x}La_{0.25+x})Cu_3O_y$ [31]. For $A_{3-x}C_{60}$ doping is introduced by 3–x cation dopants per formula unit in the type II reservoir, such that $\sigma$ is given as,

$$\sigma = \gamma \, v \, [3–x]; \text{ for } x \approx 0, \qquad (2)$$

where $v$ is the cation valence and the factor $\gamma$ derives from the allocation of the dopant by considering a given compound's structure. Equation (2) assumes that $\sigma$ is determined solely by the cation stoichiometry and that vacancy content x is randomly distributed. Following the procedure generally applied to high-$T_C$ superconductors, the charge introduced by the dopant is shared equally between the two charge reservoirs. Additionally, the methodology requires the doped charge to be distributed pair-wise between the charge-carrying layer types within each of the charge reservoirs. Consequently, $\gamma$ can be determined by applying the following two charge allocation rules [27],

(1a) Sharing between N (typically 2) ions or structural layers/surfaces introduces a factor of 1/N in $\gamma$.

(1b) Doping is shared equally between the two reservoirs, resulting in a factor of 1/2.

Given that $\gamma$ less than unity is thusly obtained, the participating charge fraction $\sigma$ is correspondingly smaller than the doping content. For optimal cuprate compounds where the doping is not known, $\sigma$ is calculated by scaling to $\sigma_0 = 0.228$ of $YBa_2Cu_3O_{6.92}$, as discussed in [27,31,32], along with the role played by electronegativity [32].

### 2.2. Application to cubic $A_3C_{60}$

The $A_3C_{60}$ compounds are treated herein as Coulomb-based high-$T_C$ superconductors, focusing on the A15 and fcc optimal forms of $Cs_3C_{60}$. Although the ambient-pressure fcc $A_3C_{60}$ superconductors have attracted the early attention in this field [22], equation (1) is not expected to accurately reflect measured values of $T_C$ for these materials because of the non-optimal behaviors pointed out earlier.

The A15 and fcc forms of $Cs_3C_{60}$ provide for an interesting case of the subject model in which the Coulombic pairing involves interactions between interfacial structures formed by the surface of the type I reservoir comprising an individual $C_{60}$ fullerene and the type II reservoir consisting of neighboring interstitial Cs cations distributed on an enclosing virtual surface. Both reservoirs each contain a single layer per formula unit, i.e. $\upsilon = 1$ and $\eta = 1$ for types I and II, respectively, in the notation of [27]. Thus the area $A$ in equation (1) is defined naturally as the surface area of the $C_{60}$ molecule, whereas the area of the virtual surface of Cs ions is inconsequential in determining $T_{C0}$. Moreover, $\zeta$, determined as below, differs significantly between the two structural forms. Given the n-type character of these materials, one associates the superconducting condensate with the $C_{60}$ molecules and with the pairing mediated through Coulomb interactions with the (presumably) positive charges on the cations.

The $C_{60}$ molecule has a diameter of 7.1 Å (radius $R = 3.55$ Å) [40],[2] and comprises 12 regular pentagons with a C–C bond length of $d_{5:6} = 1.45$ Å and 20 hexagons with C–C bond lengths of $d_{5:6}$ and $d_{6:6} = 1.40$ Å. From an equal-weighting average of $d = (1.45 + 1.40)/2 = 1.425$ Å [41], the total surface area of the $C_{60}$ molecule at standard temperature and pressure, modeled with regular polygons, is $A = 20(5.2757 \text{ Å}^2) + 12(3.6173 \text{ Å}^2) = 148.922 \text{ Å}^2$. For simplicity, $C_{60}$ area $A$ is assumed to remain essentially invariant with alkali doping and applied pressure [42]. Since $\upsilon = \eta = 1$, the $\gamma$ factor from rule (1a) is unity and that from rule (1b) is 1/2, giving $\gamma = 1/2$. From equation (2) one determines $\sigma = (1/2)[3–x]$, and equation (1) reduces to,

---

[2] $C_{60}$ diameter is given as 7.113(10) Å at ambient temperature, whence the approximation 7.1 Å derives.

$$T_{C0} = (72.28 \text{ K-Å}) [3-x]^{1/2} \zeta^{-1}, \quad (3)$$

where $x \approx 0$ is near the optimal stoichiometric value ($x = 0$ was determined for fcc $A_3C_{60}$ superconductors at ambient pressure [43]). Also following from equations (1) and (2) is the interaction charge spacing $\ell = (17.258 \text{ Å}) [3-x]^{-1/2}$.

The interaction distance $\zeta$ is found by taking the average projected radial distance between a given $C_{60}$ ion, comprising the type I structure, and its nearest neighbor Cs ions, comprising the type II structure, noticing that for both the A15 and fcc structures, the Cs on interstitial sites are facing points on $C_{60}$ unoccupied by C. From this perspective, the Cs and C positions have direct correspondence to layered structures in, e.g. intercalated β-form group-4-metal nitride-halides [35].

In the A15 structure (bcc packing, space group Pm3n), the Cs are located at tetrahedral (T) sites facing the $C_{60}$ hexagons [9]. The A15 interacting structural unit is illustrated in figure 1 (a), where the 12 Cs at (T) sites form an icosahedron enclosing the truncated icosahedron representing the $C_{60}$. The length $\zeta$ is the distance from the (T) site to the nearest C-hexagon center.

In the fcc structure (Fm3m), the Cs are distributed between (T) and octahedral (O) sites with respective occupancy ratio 2:1, and face C-hexagons or C-6:6 bonds, respectively [8]. The fcc interacting structural unit is illustrated in figure 1(b), where the 8 Cs at (T) sites and the 6 Cs at (O) sites comprise the $C_{60}$ nearest neighbors. Unlike the A15 polymorph, the $C_{60}$ molecules of the fcc form exhibit merohedral disorder [12,41], and the Cs occupying the (O) sites show considerable Debye-Waller factor disordering (0.372(1) Å rms at 30 K) [8]. The A15 and fcc structures each have 3 Cs per $C_{60}$, accounting for the 4-fold and 6-fold coordinations at (T) and (O) sites, respectively.

At ambient pressure, samples of nearly stoichiometric $Cs_3C_{60}$ are observed to be Mott insulators, exhibiting antiferromagnetism with Néel temperatures of ~46 and ~2 K for the A15 and fcc forms, respectively [7,8]. Under hydrostatic pressure, these two forms become superconducting. Optimal superconductivity for $Cs_{3-x}C_{60}$ in the A15 form occurs at applied pressure $P = 0.93$ GPa with $T_C^{meas} = 38.3$ K, $V_{per\,C_{60}} = 766.8$ Å$^3$ and lattice parameter $a_0 = 11.532$ Å, as determined from the highest $T_C$ presented in figure 4a of [8] and figure 4 of [7]; Cs stoichiometry $3-x = 2.85(1)$ is reported in

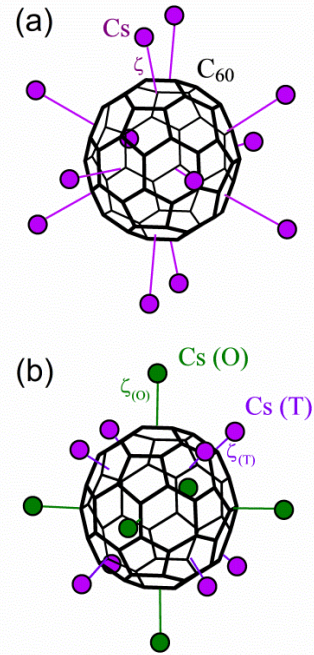

**Figure 1.** Structures of $C_{60}$ and nearest neighbor Cs sites in $Cs_3C_{60}$. (a) A15 structure, showing the 12 tetrahedral Cs sites (magenta symbols) at a distance $\zeta$ from $C_{60}$ hexagon faces; (b) fcc structure, showing the 8 tetrahedral Cs sites (magenta symbols) at a distance $\zeta_{(T)}$ from $C_{60}$ hexagon faces and the 6 octahedral Cs sites (green symbols) at a distance $\zeta_{(O)}$ from $C_{60}$ 6:6 bonds.

[9].[3] The data indicate some sample dependence, with the highest $T_C$ ranging from 38.0 to 38.3 K and the corresponding $P$ varying between 0.71 and 0.93 GPa; A15-form enrichment varies up to 77.7(6)%, the remainder being bco- and fcc-phase material [7].

For fcc $Cs_{3-x}C_{60}$, $T_C^{meas}$ = 35.2 K occurs at $P$ = 0.73 GPa with $V_{per\ C_{60}}$ = 759.6 Å$^3$ and lattice parameter $a_0$ = 14.4838 Å, as determined from the increasing-$P$ datum of highest $T_C$ in figure 3a and figure 4a of [8]. Data determining highest $T_C$ and corresponding $P$ lie in ranges 34.2–35.2 K and 0.73–0.79 GPa, respectively. The Cs stoichiometry is 3–x = 2.901(6) and fcc-form enrichment is 85.88(2)% [8].

Determining $\zeta$ for the A15 form is straight forward since Rietveld refinement indicates that the Cs cations occupy only the hexagon-coordinated (T) 6$d$ sites; the 6$c$ sites are left empty [see figure 1(a)]. Given that the (T) site is located a distance $5^{1/2}a_0/4$ = 6.4463 Å from the $C_{60}$ center, and the distance from the C-hexagon center to the $C_{60}$ center is $(R^2 - d^2)^{1/2}$ = 3.2514 Å, one has for the interaction distance $\zeta$ = $5^{1/2}a_0/4 - (R^2 - d^2)^{1/2}$ = 3.1952 Å. From this, equation (3) gives $T_{C0}$ = 38.19(7) K with $\ell$ = 10.223(18) Å for 3–x = 2.85(1) and 39.18 K with $\ell$ = 9.9639 Å upon setting $x$ = 0, results that are within 0.1 and 0.9 K, respectively, of experiment. For the fcc structure, $\zeta$ is determined by averaging the $\zeta_{(T)}$ and $\zeta_{(O)}$ distances shown in figure 1(b). From [12], the (T) sites are located over the center of the C-hexagons at the distance $3^{1/2}a_0/4$ from the $C_{60}$ center and the distance $\zeta_{(T)} = 3^{1/2}a_0/4 - [R^2 - d^2]^{1/2}$ = 3.0202 Å from the C-hexagon center. The (O) sites lie above the midpoint of C-6:6 bonds, which are at distance $a_0/2$ from the $C_{60}$ center, and the distance of $\zeta_{(O)} = a_0/2 - [R^2 - (d_{6:6}/2)^2]^{1/2}$ = 3.7616 Å from the C-6:6 bond. Using the same procedure for treating the layered high-$T_C$ materials [27] and averaging these distances over the eight (T) and six (O) Cs neighbors surrounding a given $C_{60}$, $\zeta$ = $(8/14)\zeta_{(T)}$ + $(6/14)\zeta_{(O)}$ = 3.3380 Å. From equation (3), one obtains $T_{C0}$ = 36.88(4) K with $\ell$ = 10.134(10) Å for 3–x = 2.901(6). While the agreement with experiment is reasonable, the 1.7-K difference is sufficiently large to suggest that structural disorder may be a factor. Deviation from the assumed random Cs distribution, e.g. conjecturing that the (T) sites contain 5.0 (1) % vacancies and full occupation of the (O) sites, potentially increases $\zeta$ by 0.0255 Å and reduces $T_{C0}$ by 0.28 K. Setting x = 0 in equation (3), with $\ell$ = 9.9639 Å, yields $T_{C0}$ = 37.51 K.

**Table 1.** Structural and electronic parameters of optimal A15 [7] and fcc [8] $Cs_{3-x}C_{60}$ superconductors. From experimental data for given structural form and Cs stoichiometry are optimal applied pressure $P$ with corresponding lattice parameter $a_0$, transition temperature $T_C^{meas}$ and interaction distance $\zeta$. Optimal transition temperature $T_{C0}$ is calculated for experimental x and projection to x = 0.

| Structure | 3–x | $P$ (GPa) | $a_0$ (Å) | $T_C^{meas}$ (K) | $\zeta$ (Å) | $T_{C0}$ (K) | $T_{C0}$ (x=0) (K) |
|---|---|---|---|---|---|---|---|
| A15 | 2.85(1) | 0.93 | 11.5315 | 38.3 | 3.1949 | 38.19 | 39.18 |
| fcc | 2.901(6) | 0.73 | 14.4838 | 35.2 | 3.3380 | 36.88 | 37.51 |

---

[3] Note that highest-$T_C$ pressure of 7.9 kbar is corrected higher in [7].

The relevant structural and electronic values for A15 and fcc $Cs_3C_{60}$ are listed in Table 1 and the results for $T_C^{meas}$ are graphically presented as functions of $(\ell\zeta)^{-1}$ in comparison to other high-$T_C$ compounds in figure 2; $(\ell\zeta)^{-1}$ = 0.0306 and 0.0296 Å$^{-2}$ for A15 and fcc $Cs_3C_{60}$, respectively. The diagonal line is the theoretical expression for $T_{C0}$ of equation (1). Given the relatively small variation in $\ell$, the higher $T_C^{meas}$ and $T_{C0}$ of the A15 form appear largely determined by the smaller $\zeta$ and absence of structural disorder.

## 3. Discussion

Having identified the optimal and near-optimal compounds, A15 and fcc $Cs_3C_{60}$ (at optimal $V_{per\ C_{60}}$), respectively, and accurately calculating their transition temperatures, consideration is now given to understanding the electronic origins of the pairing mechanism, citing evidence for intra-cell charge transfer and of the $e^2/\zeta$ interaction potential. Suppression of $T_C^{meas}$ below $T_{C0}$, as displayed by the fcc form, is also assessed from the perspective of merohedral and Cs(O)-site disordering, with comparisons to $Rb_xCs_{3-x}C_{60}$ as well as alkali-intercalated TiNCl superconductors, where disorder-induced pair breaking is evident. Phonon-related phenomena are discussed in the context of interlayer Coulombic pairing.

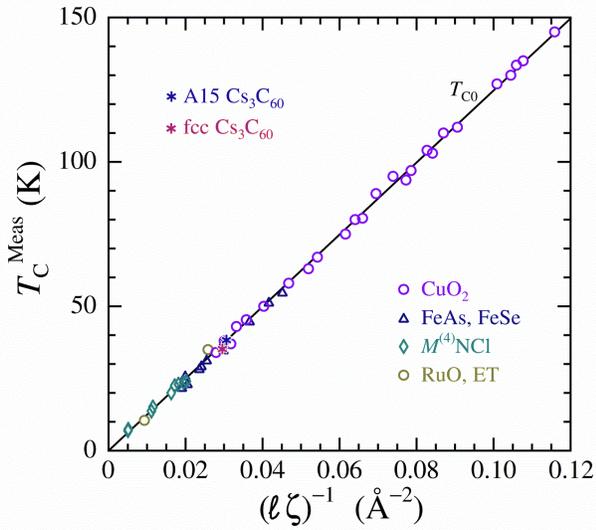

**Figure 2.** Measured optimal transition temperature $T_C^{meas}$ versus $(\ell\zeta)^{-1}$ for A15 $Cs_{2.85(1)}C_{60}$ and fcc $Cs_{2.901(6)}C_{60}$ (legend upper left), compared to other high-$T_C$ superconductors: cuprates; iron pnictides and chalcogenides; intercalated group-4-metal nitride-chlorides; and RuO- and ET-based compounds (legend lower right). The line represents $T_{C0}$ defined in equation (1).

### 3.1. A15 $Cs_3C_{60}$

Currently available experimental data on A15 $Cs_3C_{60}$ are reported for samples containing Cs vacancies along with minority phase bco and fcc inclusions [7]. Nevertheless, the A15 form stands out as a special case because of minimal disorder in Cs positions and $C_{60}$ orientations and, additionally, a unique Cs-$C_{60}$ separation determines the interaction distance $\zeta$. Agreement of $T_{C0}$ to within 0.3 % of $T_C^{meas}$ for A15-form $Cs_{2.85(1)}C_{60}$ under optimal applied pressure confers experimental validation to the Cs-$C_{60}$ Coulomb interaction model of high-$T_C$ superconductivity presented in section 2. In the following, independent experimental evidence is related to the parameters $\sigma$ and $\zeta$ in equation (1).

In allocating doping charges equally between the two reservoirs, as specified by rule 1(b), the fractional charge $\sigma$ is distributed evenly among the 12 Cs ions hosting the type II reservoir, predicting a reduction of the Cs charge state by 1/12 relative to unity. The electronic charge hosted by Cs also becomes superconducting, as shown by the gap-opening drop in NMR spin-lattice relaxation rate $(^{133}T_1T)^{-1}$ of $^{133}Cs$ for $T < T_C$ [44]. As is normally expected for metallic bonding, for example, non-integral and incomplete alkali-to-$C_{60}$ charge transfer was previously calculated for $K_3C_{60}$ from crystal orbital theory [45].

Normal-state ($T > T_C$) NMR results, $(^{133}T_1T)^{-1} = 0.0309$ s$^{-1}$K$^{-1}$ for $^{133}$Cs and $(^{13}T_1T)^{-1} = 0.0182$ s$^{-1}$K$^{-1}$ for $^{13}$C (figure 4(b) for $P = 11$ kbar in [26]), provide information on the relative electronic densities of states at the Cs and C sites, $N_{Cs}(E_F)$ and $N_C(E_F)$, respectively. Using the scaling approximation, $(^{133}T_1T)^{-1}/(^{13}T_1T)^{-1} \approx (\langle A_{Cs}^2\rangle/\langle A_C^2\rangle) [N_{Cs}(E_F)/N_C(E_F)]^2$, with hyperfine coupling constants determined for closely related fcc-form compounds, $\langle A_{Cs}^2\rangle^{1/2} = 2\pi$ 876 MHz (table S3 for Cs(T) sites in [12]) and $\langle A_C^2\rangle^{1/2} = 2\pi$ 4.36 MHz (table S4 in [13]), one obtains the ratio $N_{Cs}(E_F)/N_C(E_F) \approx 0.092$; the modeled charge on Cs relative to $C_{60}$, i.e. $1/12 = 0.083$, stands in good agreement. This result indicates that an appreciable fraction of the electronic density of states resides on the Cs ions, providing the means for the Coulombic pairing interaction.

The Coulomb potential $e^2/\zeta$ imbedded in the expression for $T_{C0}$ in equation (1) reflects transfer of unit charge between the two charge reservoirs; the possibility exists that a perturbing external electromagnetic field can induce electron transfer from a $C_{60}$ ion to one or more of its neighboring Cs ions. The externally probed charge transfer energy is screened in this case and given by $e^2/n\varepsilon\zeta$, where $\varepsilon$ is the optical dielectric constant and $n$ is the number of Cs ions participating in the charge transfer excitation. Forming an optically active electric dipole involves a portion of the 12 available nearest neighbor Cs ions, which constrains $n$ to values of about 6 or fewer. Taking $\varepsilon = 4.4$ from band structure theory [20] and $\zeta = 3.1949$ Å from Table 1, one derives local oscillator energies $\hbar\omega_n = e^2/n\varepsilon\zeta$ in the range 0.17eV to 1.02 eV for $n$ from 1 to 6. For comparison to experiment, the optical conductivity $\sigma_1(\omega)$ reported for a compressed phase of A15 Cs$_3$C$_{60}$ ($P = 18$ kbar) shows a broad distribution in the mid-infrared component, resolved over the energy range ~0.1 to ~0.84 eV (figure 1I in [46]). The peak observed at 0.17 eV and the maximum evident at ~0.33 eV are in good accord with predicted $\hbar\omega_6 \approx 0.17$ eV and $\hbar\omega_3 \approx 0.34$ eV, respectively.

Recognizing that the trend of $T_C$ increasing with $V_{per\ C_{60}}$ for fcc $A_3C_{60}$ compounds extends to optimal A15 Cs$_3$C$_{60}$, differences in electronic properties are also worthy of note. Experiment finds comparatively small values of plasma energy $\hbar\omega_p \approx 0.35$ eV and damping factor $\hbar\gamma_p \approx 0.05$ eV in compressed-phase A15 Cs$_3$C$_{60}$ (estimates from Drude-Lorentz component in $\sigma_1(\omega)$ from figure 1I in [46]), in relation to Rb$_3$C$_{60}$ (0.89 and 0.30 eV, respectively) and K$_3$C$_{60}$ (1.08 and 0.18 eV, respectively) [47]. Lowest $\omega_p$ indicates strongest Coulomb repulsion, while lowest $\gamma_p$ reflects the minimized structural disorder. When this family of compounds is compared from the perspective of band structure theory, optimal A15 Cs$_3$C$_{60}$ has the largest band width and the weakest modeled correlation strength, including expanded $V_{per\ C_{60}}$ at the metal-insulator-transition [20]. Together, these results portend diminished phononic behavior in the optimal superconducting state of A15 Cs$_3$C$_{60}$.

Spin-lattice relaxation measurements of the reduced superconducting gap for A15 Cs$_3$C$_{60}$ give values of $2\Delta/k_BT_C = 5.9$ and 4.86 for $^{133}$Cs-NMR (at 0.59 and 1.1 GPa), and 5.4 and 4.4 for $^{13}$C-NMR (at 0.58 and 1.1 GPa) [26]. Assuming the trend follows through the peak at 0.93 GPa, these results indicate strong-coupling for the optimal superconducting state.

### 3.2. fcc Cs$_3$C$_{60}$

Among the structural distinctions of optimal fcc-form Cs$_3$C$_{60}$ are that the Cs(O)-C$_{60}$ separation $\zeta_{(O)}$ is disordered by 10%, as determined from the Cs(O) Debye-Waller factor, and is on average 24.5% larger than the Cs(T)-$C_{60}$ separation $\zeta_{(T)}$ [8]. For fcc $A_3C_{60}$ generally, the interaction distance $\zeta$ is the weighted average of $\zeta_{(O)}$ and $\zeta_{(T)}$; an average of reciprocal distances may also be considered, reducing the value of $\zeta$ for optimal fcc $Cs_3C_{60}$ by 1.2%.

Comparing results in section 2 for fcc $Cs_{2.901(6)}C_{60}$ under optimal applied pressure, $T_{C0}$ exceeds $T_C^{meas}$ by 4.7(1) %. A possible origin for this divergence could be suppression of $T_C$ owing to pair-breaking caused by the intrinsic structural disorder. Since the difference between $T_{C0}$ and $T_C^{meas}$ is small, the pair-breaking parameter is approximately given as $\alpha = (4/\pi)k_B(T_{C0}-T_C^{meas}) = 0.19$ meV [48]. As expected for presumably weak perturbations on superconductive pairing, this result finds $\alpha$ is considerably smaller than the damping factors found for $A_3C_{60}$ compounds (e.g. $\hbar\gamma_p \geq 50$ meV). An analogous pair-breaking effect obeying a remote Coulomb scattering (RCS) form, $\alpha_{RCS} = x\, a_1 \exp(-k_1\zeta)$, was previously shown to fit data on alkali-intercalated TiNCl superconductors with interaction distance $\zeta$, alkali content $x$ per TiNCl, and fitted parameters $a_1 = 23.9(1.0)$ meV and $k_1 = 0.727(23)$ Å$^{-1}$ [34]. Applying this result to the Cs(O) sites, using $\zeta_{(O)}$ and taking $x = 6/60 = 0.1$, a prediction $\alpha_{RCS} = 0.16$ meV is obtained. That $\alpha_{RCS}$ is close to $\alpha$ calculated from $T_C^{meas}$ suggests that Cs(O) disorder is effective in suppressing the transition temperature to the level observed.

It is instructive to compare fcc $Cs_3C_{60}$ with the fcc $Rb_xCs_{3-x}C_{60}$ alloys under applied pressures, for which $T_C$ vs. $V_{per\,C_{60}}$ exhibits maxima $T_C^{max}$. These maxima and the corresponding $V_{per\,C_{60}}$ systematically decrease with x, as shown in figure 3. Data for x = 0 are for the optimal fcc $Cs_{2.901(6)}C_{60}$ [8]; data for $0.35 \leq x \leq 1$ are read from figure 1(E) in [13], which indicates little change in $T_C^{max}$ for $1 \leq x \leq 2$.

The decrease in $T_C^{max}$ with x suggests an additional pair breaking effect is contributed by alloying disorder, while the in decrease $V_{per\,C_{60}}$ is likely associated with the decrease in average cationic size. Considering the results for $RbCs_2C_{60}$, one can explain the suppressed $T_C^{max} = 33.2$ K at $V_{per\,C_{60}} = 755.5$ Å$^3$ with pair breaking parameter $\alpha = 0.49$ meV and theoretical $T_{C0} = 37.63$ K ($\zeta = 3.3259$ Å, $\ell = 9.964$ Å).

The transformation from weak-coupling BCS-like behavior to unconventional superconductivity in fcc $Cs_3C_{60}$ may be illustrated by considering NMR and specific heat data for the $Rb_xCs_{3-x}C_{60}$ alloys at ambient pressure. For $Rb_3C_{60}$ at x = 3, for which $V_{per\,C_{60}} = 737$ Å$^3$ is under expanded relative to optimal $Cs_3C_{60}$, $2\Delta/k_BT_C = 3.6(1)$, determined from $^{13}$C NMR $T_1$ measurements, and the specific heat jump at $T_C$ are both consistent with weak-coupling BCS theory [13]; weak coupling is reported for $K_3C_{60}$ [13] and several of the ambient-pressure fcc $A_3C_{60}$ superconductors as well [19]. The optimally compressed $Cs_3C_{60}$ at $V_{per\,C_{60}}= 759.6$ Å$^3$ is bounded on either side by two ambient-pressure compounds; $RbCs_2C_{60}$ with $V_{per\,C_{60}} = 762$ Å$^3$ is slightly over expanded and $Rb_2CsC_{60}$ with $V_{per\,C_{60}} = 741$ Å$^3$ is under expanded, both relative to optimal. For $RbCs_2C_{60}$, where pressure dependence data indicate closest proximity to $T_C^{max}$ at ambient pressure, $2\Delta/k_BT_C = 4.9$ and the specific heat jump at $T_C$ is also strongly enhanced above weak-coupling BCS [13,26]; in comparison, $2\Delta/k_BT_C = 4.3$ for $Rb_2CsC_{60}$, with a correspondingly smaller, but still significant, enhancement in the specific heat jump [13]. These results are also confirmed by $^{87}$Rb NMR. Differing trends are observed for fcc $Cs_3C_{60}$ under applied pressures, where $2\Delta/k_BT_C = 4.3(1)$ at $V_{per\,C_{60}} = 766$ Å$^3$ from $^{133}$Cs NMR, indicating strong coupling, and $2\Delta/k_BT_C = 3.4(1)$ and $3.8(2)$ at $V_{per\,C_{60}} = 757$ and $762$ Å$^3$, respectively, indicating weak coupling, as read from figure 3(c) in [49]. Another inconsistency is found near the metal-insulator-transition, where NMR $T_1$ indicates strong coupling for fcc $Cs_3C_{60}$ and $Rb_{0.35}Cs_{2.65}C_{60}$, whereas weak coupling agrees with the specific heat jump in the alloy [13,49]. While findings of strong coupling are noted hallmarks of high-$T_C$ superconductivity [50], such discrepancies remain to be explained.

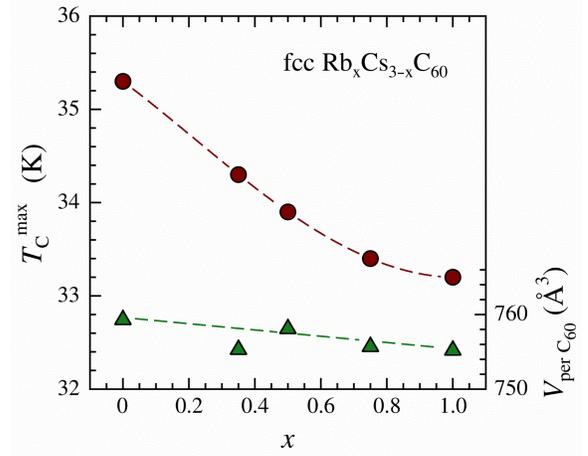

**Figure 3.** Maximum measured transition temperature $T_C^{max}$ plotted against Rb content x in $Rb_xCs_{3-x}C_{60}$ for x = 0 (from [8]) and x = 0.35, 0.5, 0.75 and 1 (from [13]), denoted by circle symbols. Triangles denote corresponding measured $V_{per\,C_{60}}$ (from [8] and [13]). Dashed curves are guides to the eye.

### 3.3. Origin of Phononic Effects

For ambient-pressure fcc $A_3C_{60}$ superconductors, results drawn from isotopic substitution, electronic specific heat, NMR, lattice expansion and other experimental techniques have provided much impetus for basing pairing theory on $C_{60}$-localized vibrations, molecular distortion, on-site Coulomb repulsion $U$ and $t_{1u}$ band width $W$; using viable estimates of electron-phonon coupling $\lambda = 0.3$, $U/W = 1.5$ and $W = 0.6$ eV, however, places a limit of 15–20 K on the highest calculated BCS-like $T_C$ [19,22].

Comparison with optimally compressed forms of $Cs_3C_{60}$, exhibiting unconventional non-BCS-like signatures and significantly higher $T_C$ values, suggests that the ambient-pressure fcc $A_3C_{60}$ compounds are formed with non-optimal $V_{per\,C_{60}}$. It is also clear that interlayer $C_{60}$-Cs Coulomb mediation dominates in $Cs_3C_{60}$ at optimal $V_{per\,C_{60}}$, as revealed by the excellent agreement between theory and experiment.

In addition to the dome in $T_C$ versus $V_{per\,C_{60}}$, evidence supporting an underlying novel superconductivity mechanism extending even into the non-optimal regime include maximum $T_C$ occurring at stoichiometric doping [18], absence of resistivity saturation at high temperature [51], transformation in resistivity from a high-order temperature dependence for $K_3C_{60}$ to a linear temperature dependence for $Rb_3C_{60}$ at constant volume [52] and the presence of a spin gap below 100 K [8], even though some arguments have been put forth in terms of phononic pairing [22]. Consequently, an intriguing question concerning the non-optimal superconducting states is whether a portion of the Coulombic pairing energy is imparted to $C_{60}$-ion phononics, inducing some BCS-like phenomena. In this view of materials characterized by $T_C \ll T_{C0}$, intra-$C_{60}$ dynamics provide a sympathetic response to the native Coulombic-based superconductivity.

Sympathetic response of the lattice may also account for the appearance of a weak-coupling BCS signature in under compressed material near the insulating phase boundary, as noted above [13]. This is the region of the $T_C$-$V_{per\,C_{60}}$ phase diagram where a strong-correlation Hubbard model for fcc $Cs_3C_{60}$ with negative Hund's coupling finds strongest phonon involvement [23-25]. In the vicinity of optimal $V_{per\,C_{60}} \approx 762$ Å$^3$, this theory provides the result $T_C \approx 20$ K (read from figure 2 in [23]) falling significantly below the experimental $T_C^{meas}$ = 35.2 K. Given the limitations on $T_C$ obtained from two phonon-based models, the influence of phonon-based interactions appears mainly confined to phases of non-optimal $V_{per\ C_{60}}$.

## 4. Conclusion

It is argued that the A15 and fcc forms of $Cs_3C_{60}$, with transition temperatures maximized under hydrostatic pressure, respectively represent the optimal and near-optimal superconducting compounds of these two representative macrostructures of the $A_3C_{60}$ family. This designation is supported by previously established phase diagrams and other experimental evidence.

The agreement of $T_{C0}$ (=38.19 K) to within 0.3 % of $T_C^{meas}$ (=38.3 K) for A15 $Cs_{2.85(1)}C_{60}$ at optimal $V_{per\ C_{60}}$ confirms the high-$T_C$ nature of the optimal superconducting state and the validity of the Cs-$C_{60}$ Coulomb pairing interaction as described in section 2. By analyzing NMR $(^{133}T_1T)^{-1}$ and $(^{13}T_1T)^{-1}$ measurements of A15 $Cs_3C_{60}$, in combination with hyperfine coupling constants derived for closely-related fcc compounds, an estimate of the ratio $N_{Cs}(E_F)/N_C(E_F) \sim 9\%$ is derived, which is similar in magnitude as the modeled 1/12 (~8%) charge allocation on Cs relative to $C_{60}$. Additionally, it is shown that evidence of the Coulomb potential $e^2/\zeta$, imbedded in equation (1), is found in the broad mid-infrared component of the optical conductivity of A15 $Cs_3C_{60}$ [46], which is modeled by optically active electric dipoles of local oscillator energies $\hbar\omega_n = e^2/n\varepsilon\zeta$ forming between the $C_{60}$ and a portion $n$ of the 12 nearest-neighbor Cs cations. In particular, $\hbar\omega_6 \approx 0.17$ eV and $\hbar\omega_3 \approx 0.34$ eV are in excellent agreement with the peak observed at 0.17 eV and the maximum evident at ~0.33 eV, respectively.

For fcc $Cs_{2.901(6)}C_{60}$ under optimal applied pressure, $T_{C0}$ is seen to exceed $T_C^{meas}$ by 1.7 K, which is attributed to the suppression of $T_C$ owing to disorder-induced pair-breaking with $\alpha \approx (4/\pi)k_B(T_{C0}-T_C^{meas}) = 0.19$ meV [48]. Drawing an analogy with RCS pair breaking in alkali-intercalated TiNCl [34], and applying this result to the Cs(O) sites, a value of $\alpha_{RCS} = 0.16$ meV is obtained in reasonable agreement with $\alpha$ above. This agreement indicates that the Cs(O)-site fluctuations may be responsible for the observed suppression in $T_C^{meas}$. Extending the calculations of $T_{C0}$ to x = 0, assuming unchanged structural parameters, yields values of 39.18 and 37.51 K for the A15 and fcc forms, respectively.

For non-optimal $A_3C_{60}$ materials, a hypothesis is presented attributing the observed BCS-like phenomena to a sympathetic response of the lattice to the native Coulomb-mediated superconductivity. This deduction follows from inadequately low values of $T_C$ and absences of either $T_C$-$V_{per\ C_{60}}$ or stoichiometry optimization domes evident in pairing theories based on $C_{60}$ phononics, as well as other theoretical shortcomings that have been noted [19,22]. With their focus on alkali $A$ for chiefly controlling doping and lattice spacing, prior theoretical works have notably overlooked the electronic charges at $A$ sites that are essential for high-$T_C$ superconductivity.

In conclusion, the successful predictions given herein demonstrate the validity of the interfacial Coulombic pairing theory as applied to the superconductivity in optimal $Cs_3C_{60}$ and the near-optimal $A_3C_{60}$ family in general.

## Acknowledgments

The authors are grateful for support from the College of William and Mary, the New Jersey

Institute of Technology, and the University of Notre Dame.